\documentclass[sigconf,nonacm]{acmart}  

\AtBeginDocument{%
  }

\setcopyright{none}
\renewcommand\footnotetextcopyrightpermission[1]{} 

\usepackage{multirow} 
\usepackage{tabularx}
\usepackage{balance}
\usepackage{booktabs}
\usepackage{array}
\usepackage{xcolor}
\usepackage{fixltx2e} 
\usepackage{makecell}
\usepackage{diagbox}
\usepackage{fancyhdr}

\newcommand{\mmgensrinfo}{MMGenSR '25, November 14, 2025, Seoul, Republic of Korea}

\newcommand{\runtitle}{Enhancing Medical Cross-Modal Hashing Retrieval using Dropout-Voting Mixture-of-Experts Fusion}

\fancypagestyle{mmgensr}{%
  \fancyhf{}

  \fancyhead[LE]{\footnotesize\itshape \mmgensrinfo}
  \fancyhead[RE]{\footnotesize\itshape \runauthors}
  \fancyhead[LO]{\footnotesize\itshape \runtitle}
  \fancyhead[RO]{\footnotesize\itshape \mmgensrinfo}
  \fancyfoot[C]{}
}

\fancypagestyle{mmgensr-first}{%
  \fancyhf{}

  \fancyfoot[L]{\footnotesize \mmgensrinfo}
  \fancyfoot[C]{}
}

\begin{document}
\pagestyle{mmgensr}

\title{Enhancing Medical Cross-Modal Hashing Retrieval using Dropout-Voting Mixture-of-Experts Fusion}

\author{Jaewon Ahn}
\authornote{Both authors contributed equally to this research.}
\email{vpersie@yonsei.ac.kr} 
\affiliation{%
  \institution{Graduate School of Integrative Medicine, Yonsei University}
  \city{Seoul}
  \country{South Korea}
}

\author{Woosung Jang}
\authornotemark[1]
\email{jwill1994@yonsei.ac.kr}
\affiliation{%
  \institution{Graduate School of Integrative Medicine, Yonsei University}
  \city{Seoul}
  \country{South Korea}
}

\author{Beakcheol Jang}
\authornote{Corresponding author.}
\email{bjang@yonsei.ac.kr}
\affiliation{%
  \institution{Graduate School of Information, Yonsei University}
  \city{Seoul}
  \country{South Korea}
}

\newcommand{\runauthors}{Jaewon Ahn, Woosung Jang, and Beakcheol Jang}

\renewcommand{\shortauthors}{ooo et al.}

\begin{abstract}
In recent years, cross-modal retrieval using images and text has become an active area of research, especially in the medical domain. The abundance of data in various modalities in this field has led to a growing importance of cross-modal retrieval for efficient image interpretation, data-driven diagnostic support, and medical education. In the context of the increasing integration of distributed medical data across healthcare facilities with the objective of enhancing interoperability, it is imperative to optimize the performance of retrieval systems in terms of the speed, memory efficiency, and accuracy of the retrieved data. This necessity arises in response to the substantial surge in data volume that characterizes contemporary medical practices.
In this study, we propose a novel framework that incorporates dropout voting and mixture-of-experts (MoE) based contrastive fusion modules into a CLIP-based cross-modal hashing retrieval structure. We also propose the application of hybrid loss. So we now call our model MCMFH which is a medical cross-modal fusion hashing retrieval. Our method enables the simultaneous achievement of high accuracy and fast retrieval speed in low-memory environments. The model is demonstrated through experiments on radiological and non-radiological medical datasets.
\end{abstract}




\maketitle
\thispagestyle{mmgensr-first}
\pagestyle{mmgensr}   

\begin{figure*}[t]
    \centering
    \includegraphics[width=\textwidth]{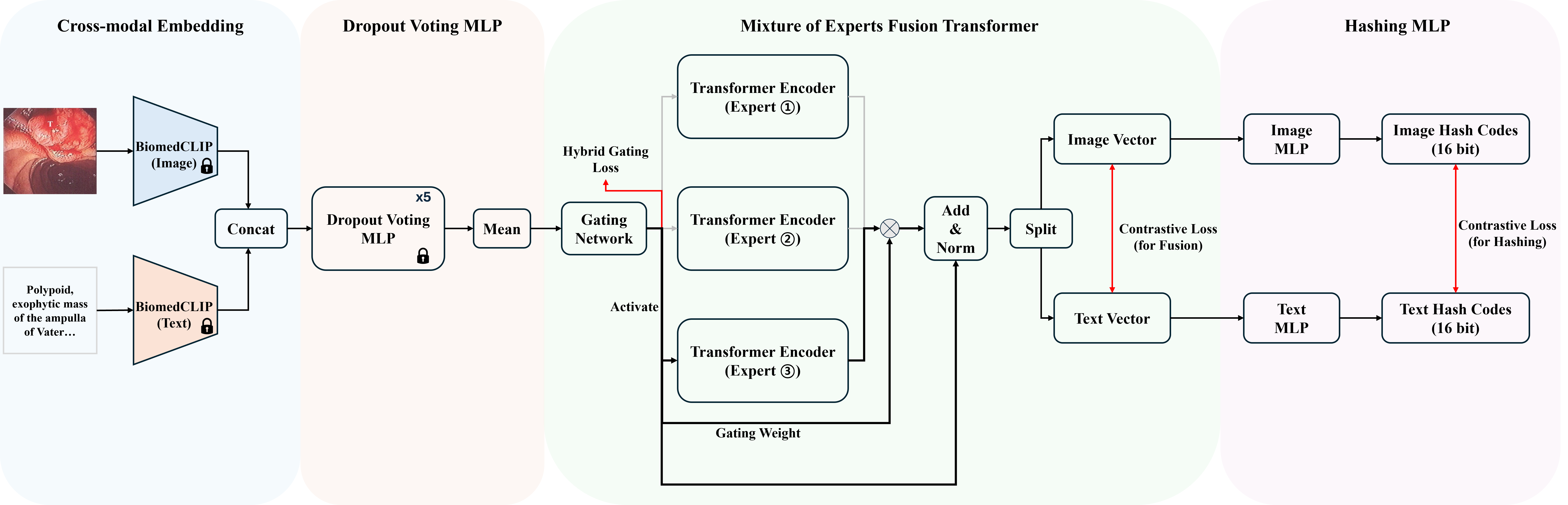}
    \Description{The overview of our proposed model MCMFH.}
    \caption{The overview of our proposed model MCMFH.}
    \label{fig:framework}
\end{figure*}

\section{Introduction}
In recent years, multimodal retrieval has emerged as a key technology in various fields such as e-commerce, social media, law, and healthcare. In particular, in the medical field, multimodal data retrieval technology is becoming more important as diagnosis and decision-making are based on various types of data such as images, text, and test results. Medical data is characterized by large volumes, multiple modalities, and unstructured data, which are major obstacles to the practical application of existing AI systems. Therefore, highly efficient and accurate multimodal search techniques can be very useful for real-time decision support in clinical settings \cite{esteva2019guide, acosta2022multimodal}.

In this work, we specifically address the most commonly studied image-to-text cross-modal retrieval problem, namely cross-modal hashing retrieval (hereafter referred to as CMHR). It maps data across different modalities into the same Hamming space, which enables efficient similarity-based retrieval. Hashing-based retrieval methods have significant advantages in terms of memory efficiency and computational speed, making them particularly suitable for clinical environments where real-time processing is required. Existing CMHR techniques can be broadly categorized into the following based on their training methods: Shallow methods, Deep hashing methods, Semi-deep methods. Most of the existing research emphasizes on speed and storage efficiency, which leads to an imbalance problem with accuracy \cite{wang2017survey, jiang2017deep, wang2025cross}.

To address these issues, various attempts have been made in recent years to improve accuracy. For example, Modality-specific Projection Learning, Semantic Consistency Preservation, and Modality-invariant Hash Code Learning via Adversarial Training have been introduced \cite{zhang2018deep, zhen2019deep, wu2020modality}. However, they often come with drawbacks such as increased model complexity and poor generalization ability.

Recently, CLIP-based models have gained attention for their strong cross-modal representation learning performance in general domains. In particular, domain-specific pre-trained models, such as BiomedCLIP, are good at semantic mapping between medical images and medical documents, and show strong performance even in a zero-shot setting. In this regard, CLIP-based models are highly advantageous to be utilized as the backbone of CMHR. Several approaches have been explored in prior work, including UCMFH, DScPH and DDBH \cite{XIA2023101968, tu2022differentiable, liu2023multi, huo2023deep, huo2024deep, huo2024deep2, tu2024two, tu2025cross, qin2025deep, zhu2025deep, qin2025deep2}.

However, existing CLIP-based CMHR methodologies mostly follow a modality-specific separation structure, and there is a lack of research with explicit fusion structures. In contrast, in the general multimodal learning field, various fusion structures (e.g., Attention-based, Transformer-based, etc.) are used and have shown good performance
\cite{lu2019vilbert, tan2019lxmert, chen2020uniter, kiela2019mmbt, tsai2019multimodal}.

The first study to introduce the Fusion module in CMHR is UCMFH \cite{XIA2023101968}. In this study, we propose a novel fusion module by integrating several methodological advances, aiming to enhance performance. More specifically, a novel form of fusion strategy that combines Dropout Voting and MoE are included to improve the performance of CMHR in the medical field \cite{tsai2019multimodal, shazeer2017outrageously, srivastava2014dropout}. The proposed structure focuses on improving accuracy while maintaining the speed and memory efficiency of hashing techniques, and shows promise in real-world clinical settings. Our main methodological contributions are as follows.

\begin{itemize}
\item{\texttt{MoE Fusion Transformer}}: For the first time, to the best of our knowledge, we have applied the Fusion Module with MoE structure to CMHR. A MoE transformer strengthens image–text fusion under the low memory budget, improving robustness on both radiology and non-radiology datasets.
\item{\texttt{Frozen Dropout Voting}}: A stochastic voting module perturbs input features extracted from the CLIP backbone and averages them, acting as a representation-level ensemble that boosts generalization without extra trainable parameters.
\item{\texttt{Hybrid Gating Loss}}: We introduce a task + load-balancing hybrid loss that regularizes the MoE gate and yields the best average performance across datasets.
\end{itemize}

\section{Method}
In this section, we will describe the proposed MCMFH model. The overall model structure can be seen in figure 1.

\subsection{Cross-modal Embedding}
To provide medical-domain-aware multi-modal features, we adopt not the original CLIP models optimized for general domains (e.g., web image-caption pairs) but BiomedCLIP \cite{zhang2025biomedclipmultimodalbiomedicalfoundation}. Trained on millions of paired medical images and reports, BiomedCLIP captures lesion cues and specialized medical terminology that standard CLIP encoders overlook. BiomedCLIP maps each medical image (\begin{math}x_{\text{image}}\end{math}) and its medical description (\begin{math}x_{\text{text}}\end{math}) into 512-dimensional embeddings, and we concatenate the vectors of these two modalities as input to a downstream model. This pre-training based medical multi-modal embedding serves as an important preprocessing step, ensuring that the hashing module is optimized for retrieval while preserving domain-specific semantics.
\begin{equation}
z = \operatorname{Concat}\left[\mathrm{BiomedCLIP}(x_{\text{image}}),\ \mathrm{BiomedCLIP}(x_{\text{text}})\right]
\end{equation}

\subsection{Dropout Voting MLP}
We prepend a two-layer MLP in front of the MoE gate.
\begin{equation}
f(z) = W_2 \, \sigma\left(\operatorname{Dropout}_{0.2}(W_1 z)\right), \quad \sigma = \mathrm{GELU}
\end{equation}
The weights \begin{math}W_{\text{1}}\end{math}, \begin{math}W_{\text{2}}\end{math} are frozen and dropout (\begin{math}p=0.2\end{math}) is kept active during both training and inference. For each sample we draw (\begin{math}K=5\end{math}) independent dropout masks.
\begin{equation}
\mathbf{z}_k = f^{(k)}(\mathbf{z}), \quad k = 1 \ldots K, \quad x = \frac{1}{K} \sum_{k=1}^{K} \mathbf{z}_k,
\end{equation}
and forward the averaged feature \begin{math}x\end{math} to the MoE Fusion Transformer (\ref{sec:MoE Fusion Transformer}).
The effect on the Dropout Voting MLP is as follows.
To enhance the robustness of the representation, we apply the dropout multiple times (K) and average the resulting K embeddings. This stochastic ensembling process suppresses noise and reinforces consistent signals aligned with the ground truth, leading to representations that are more likely to reside near the correct class; perturbation-induced correction \cite{kim2023improving}. Each forward pass activates different sub-networks, enabling the model to learn over a neighborhood rather than a single point in representation space, which improves generalization and reduces overfitting.
Additionally, the Voting MLP remains frozen during training, preventing the backbone from adapting to fixed dropout-induced perturbations, thus maintaining a consistent regularization effect \cite{zhang2024fine}. The resulting variance from dropout also encourages the gating network to generate broader routing distributions, mitigating expert collapse, and promoting balanced expert utilization.

\subsection{MoE Fusion Transformer}\label{sec:MoE Fusion Transformer}
\subsubsection{MoE Structure}
The MoE Fusion Transformer consists of three main components.
\\\textit{Gating Network}. For an input vector x, the gating network computes the selection probability of each expert via softmax.
\begin{equation}
g(x) = \mathrm{softmax}(W_{\text{gate}} x)
\end{equation}
Only the top-1 expert with the highest probability is then activated, and the rest of the experts are ignored \cite{fedus2022reviewsparseexpertmodels}.
\\\textit{Experts (Transformer Encoder)}. Each expert has two layers: a transformer encoder layer with a hidden dimension of 1024 and a self-attention based representation learning with four multi-head attention \cite{vaswani2023attentionneed, XIA2023101968}.
\\\textit{Mixture Output \& Residual Normalization}. The output of the selected expert is multiplied by the selection probability of that expert, which reflects the confidence of the gating.
\begin{equation}
h(x) = g(x) \cdot \mathrm{Expert}(x)
\end{equation}
We then applied a residual connection with input x and layer normalization.
\begin{equation}
z = \mathrm{LayerNorm}(h(x) + x)
\end{equation}
This structure allows the model to extract information from the perspectives of different experts on the same input and dynamically select the output of the most appropriate expert through a gating network to enhance inter-modal interaction.
\subsubsection{Hybrid Gating Loss}
MoE Fusion Transformer uses a combination of the following two losses to improve the routing performance of the gating network for overall learning stability.
\\\textit{Switch load-balancing loss}. Minimizing (\begin{math}\mathcal{L}_{\text{switch}}\end{math}) aligns the realized traffic (\begin{math}T_i\end{math}) with the gate distribution (\begin{math}P_i\end{math}) and discourages any single expert from monopolizing the workload, thereby balancing the MoE \cite{fedus2022switchtransformersscalingtrillion}. The specific loss function is defined as follows:
\begin{equation}
\mathcal{L}_{\text{switch}} = \lambda N \sum_{i=1}^{N} T_i \, P_i,
\end{equation}
where $N$ is the number of experts, $B$ is the batch size, and $\lambda (= 10^{-2})$ is the weighting factor.

\begin{equation}
T_i = \frac{1}{B} \sum_{n=1}^{B} r_{n,i}, \qquad
P_i = \frac{1}{B} \sum_{n=1}^{B} s_{n,i}
\end{equation}

\[
r_{n,i} = 
\begin{cases}
1 & \text{if sample } n \text{ is routed to expert } i, \\
0 & \text{otherwise}
\end{cases}
\]
\[
s_{n,i} = \text{softmax score for expert } i \text{ before top-1}
\]
\textit{Variance-based load-balancing loss}. Minimizing the (\begin{math}\mathcal{L}_{\text{var}}\end{math}), we directly reduce the variance of the probability of choosing a particular expert, which prevents the gating network from producing biased probabilities \cite{shazeer2017outrageouslylargeneuralnetworks}. The specific loss function is defined as follows:
\begin{equation}
\mathcal{L}_{\text{var}} = N \sum_{i=1}^{N} \left( p_i - \frac{1}{N} \right)^2
\end{equation}
\\\textit{Hybrid Gating Loss}. We configured it to compute a hybrid loss as a weighted sum of the two losses. This allows the gating network to reduce the bias of the expert selection probability and keep the actual selection distribution uniform, thus ensuring the stability of inter-expert learning.

\subsubsection{Fusion Contrastive Learning}
The vector z transformed by the MoE Fusion Transformer is split into image vector \( \mathbf{z}^v \) and text vector \( \mathbf{z}^t \) in the same order as when it was originally concatenated. We apply the same fusion contrastive loss (\begin{math}\mathcal{L}_{\text{fusion}}\end{math}) used in UCMFH \cite{XIA2023101968} to induce cross-modal alignment. This loss maximizes the similarity between aligned image-text pairs and trains them to be distinct from the rest of the pairs.

\subsection{Hashing MLP}
Following UCMFH \cite{XIA2023101968}, we reuse the same two-layer hashing MLP (ReLU–Dropout–Tanh) and \textit{sign} quantization to map the fused image and text features to hash codes. The subsequent hash contrastive loss (\begin{math}\mathcal{L}_{\text{hash}}\end{math}) is identical to that work, so we refer readers to UCMFH for implementation details.

\subsection{Overall Objective Function}
The overall objective function is organized as follows The weighting values for each element were specified through experimentation.
\begin{equation}
\mathcal{L} = \mathcal{L}_{\text{fusion}} + 0.85 \cdot \mathcal{L}_{\text{switch}} + 0.15 \cdot \mathcal{L}_{\text{var}} + 0.5 \cdot \mathcal{L}_{\text{hash}}
\end{equation}

\begin{table*}
  \caption{Comparative Analysis of Baselines, Proposed Model, and Ablation Variants based on mAP}
  \label{tab:table_1}
  \begin{tabularx}{\textwidth}{c l *{6}{>{\centering\arraybackslash}X}}
    \toprule 
    \textbf{Exp.} & \textbf{Method\ \textbackslash\ Dataset} 
      & \multicolumn{3}{c}{\textbf{open-i}} 
      & \multicolumn{3}{c}{\textbf{roco/non-rad}} \\
    \cmidrule(lr){3-5} \cmidrule(lr){6-8}
    & & \texttt{I2T-16 (32/64)} & \texttt{T2I-16 (32/64)} & \texttt{Mean-16 (32/64)} 
      & \texttt{I2T-16 (32/64)} & \texttt{T2I-16 (32/64)} & \texttt{Mean-16 (32/64)} \\
    \midrule
    \multirow{4}{*}{\makecell{\textbf{Main}\\\textbf{Comp.}}}
      & \texttt{DScPH} & .194(.247/.280) & .186(.208/.235) & .190(.227/.257) & .526(.525/.551) & .516(.516/.520) & .521(.520/.535) \\
      & \texttt{DDBH} & .401(.432/\textbf{.449}) & .297(.299/.318) & .349(.366/.384) & .564(.598/.600) & .538(.565/.566) & .551(.582/.583) \\
      & \texttt{UCMFH} & .354(.367/.382) & .371(.375/.391) & .362(.373/.386) & .634(.657/.646) & .653(\textbf{.661}/.659) & .643(\textbf{.659}/.652) \\
      & \texttt{MCMFH(Ours)} & \textbf{.429}(\textbf{.436}/.434) & \textbf{.423}(\textbf{.429}/\textbf{.434}) & \textbf{.426}(\textbf{.432}/\textbf{.434}) & \textbf{.687}(\textbf{.662}/\textbf{.665}) & \textbf{.699}(.652/\textbf{.674}) & \textbf{.693}(.657/\textbf{.669}) \\
    \midrule
    \multirow{5}{*}{\makecell{\textbf{Module}\\\textbf{Abl.}}}
      & \texttt{Base(UCMFH)} & .354 & .371 & .362 & .634 & .653 & .643 \\
      & \texttt{+MoE} & .412 & .412 & .412 & .655 & .628 & .641 \\
      & \texttt{+MoE+Voting1+Frozen} & .383 & .381 & .382 & .645 & .651 & .648 \\
      & \texttt{+MoE+Voting5+Unfrozen} & .404 & .395 & .399 & .650 & .649 & .649 \\
      & \texttt{+MoE+Voting5+Frozen(Ours)} & \textbf{.429} & \textbf{.423} & \textbf{.426} & \textbf{.687} & \textbf{.699} & \textbf{.693} \\
    \midrule
    \multirow{3}{*}{\makecell{\textbf{Loss}\\\textbf{Abl.}}}
      & \texttt{MCMFH w/Switch} & .421 & .410 & .416 & .680 & .693 & .687 \\
      & \texttt{MCMFH w/Variance-based} & .419 & .418 & .419 & .682 & .685 & .684 \\
      & \texttt{MCMFH w/Hybrid} & \textbf{.429} & \textbf{.423} & \textbf{.426} & \textbf{.687} & \textbf{.699} & \textbf{.693} \\
    \bottomrule
  \end{tabularx}
\end{table*}

\section{Experiments}
\subsection{Datasets and Metrics}
As a dataset to demonstrate the effectiveness of our methodology, we used the open-i data, which is one of the frequently used radiology data in the field of medical vision. \cite{openi_dataset} Meanwhile, for non-radiology data, we used the non-radiology sub-dataset of the ROCO dataset. \cite{pelka2018roco}
\begin{itemize}
\item{\textit{Open-i}}: There are a total of 121 problem types. Rare problems with less than 20 occurrences were not utilized, and only half of ‘normal’ problems with the most occurrences were kept to reduce class imbalance. Finally, we removed problems that do not correspond to diseases, such as ‘Medical Device’ and ‘No Indexing’ problems, and used 39 types in total.
\item{\textit{ROCO/Non-Radiological}}: ROCO is a multimodal medical image-to-text dataset that provides medical images collected from articles in PubMed Central, along with captions, labels (UMLS semantic types), and section information. \cite{bodenreider2004unified} We used the top 10 types of occurences based on semantic types.
\end{itemize}

In both datasets, we extracted label data based on one-hot encoding. In addition, the ratio of query: retrieval: train data was 1:6:3. To evaluate retrieval performance, we use mean Average Precision (mAP), the standard metric in CMHR, for both image-to-text (I2T) and text-to-image (T2I) tasks. Following UCMFH \cite{XIA2023101968}, we compute mAP exactly as specified in their protocol.
\subsection{Experimental Setup}
The main hyperparameters were set as follows: the batch size was 32, and the number of training epochs was 150. The learning rate was set to 0.0001 for the MoE Fusion Transformer and 0.001 for both the image and text branches of the Hashing MLP. The hash code length was primarily fixed to 16 bits, as our focus was on achieving high retrieval performance with reduced memory usage, which is crucial for practical applications in the medical domain. The Adam \cite{kingma2017adammethodstochasticoptimization} optimizer was employed for all training components.
\subsection{Performance}\label{sec:performance}
To prove its effectiveness in terms of clinical practicality, we compared the retrieval performance based on 16 bit hash code, which is low memory. In addition, as our methodology is CLIP Based, we compared UCMFH \cite{XIA2023101968} with two additional most recent CLIP Based CMHR models: DScPH \cite{10855579}, DDBH \cite{11003934}.
For a fair comparison, we unified the backbone CLIP model to bioMedCLIP and matched the hyperparameters such as epoch, batch size, and random seed. Despite DScPH and DDBH training up to the CLIP backbone, the results in Table ~\ref{tab:table_1} show that our model's mAP is larger than DScPH by 0.236 in open-i and 0.172 in roco/non-rad, and our model is larger than DDBH by 0.077 in open-i and 0.142 in roco/non-rad. We can also see that our methodology is more efficient in terms of both resources required for training and performance since we do not train CLIP itself.

Finally, we can see that the average mAP value is 0.064 larger in open-i and 0.05 larger in roco/non-rad than UCMFH, which performs rather well compared to other modern models. Additional 32 bit and 64 bit experiments also showed mostly improved performance. (Table ~\ref{tab:table_1}) This confirmed the practicality of our model, which performed well at the lowest 16 bits.

\subsection{Ablation Study}
\subsubsection{Effect of MoE}
Table ~\ref{tab:table_1} shows the effectiveness of one of our key ideas, the MoE Fusion Transformer. In particular, for open-i, we can see that it leads to a significant increase in the resulting average mAP of 0.05. For roco/non-rad, we can see that the combination of MoE and Frozen Dropout Voting MLP with expert balancing works well.

\subsubsection{Effect of Dropout Voting MLP}
The effect of Frozen on the Dropout Voting MLP can also be seen from the results of the Table ~\ref{tab:table_1} experiment. First, we can see the effect of Frozen in maximizing the dropout effect, as the average mAP value is 0.027 higher in open-i and 0.044 higher in roco/non-rad. Finally, the effect of voting can also be seen as the average mAP value is 0.044 larger in open-i and 0.045 larger in roco/non-rad when the number of votes is 5 than when the number of votes is 1. Taken together, these results show that the features input to the MoE Fusion Transformer after the Frozen Dropout Voting MLP are transformed into more generalized and stable features, which in turn leads to better performance of the overall model.

\subsubsection{Effect of Hybrid Gating Loss}
Table ~\ref{tab:table_1} shows that the hybrid loss in the final architecture has the best mAP in the entire dataset. This shows that the hybrid loss aligns the selection probability with the actual expert selection rate through the switch loss, and minimizes the variance through the variance loss. This further equalizes the selection probability and contributes to achieving stable performance improvement.

\subsubsection{Retrieval Efficiency}
Finally, to verify clinical practicality, we compared retrieval latency between 16 bit hash codes and real-valued features. On average across the two datasets, real-valued retrieval was 1.73 times slower. It is also \begin{math}1\!/\!4\end{math} more memory efficient than 64 bit, which increases its practicality in terms of speed and memory efficiency.

\section{Conclusion}
In this paper, we propose MCMFH on the CMHR task and take a step towards its practicality in the clinical environment by demonstrating the balancing effectiveness of MCMFH in the medical domain, which satisfies both resource efficiency and mAP performance. As shown in the experimental results, MCMFH outperforms the existing baseline and recent CLIP-based CMHR methods, DScPH and DDBH, in both I2T and T2I at a lower bit count of 16 bits. Future work will extend beyond cross-modal retrieval to evaluate the effectiveness of the proposed approach in environments involving three or more modalities, using larger and more diverse medical datasets.


\balance
\section*{GenAI Usage Disclosure}
Generative AI tools were used exclusively for minor editing, grammar correction, and providing ideas specifically for the outline of the manuscript, as well as for translating portions of the text. All scientific content, analysis, and interpretation are entirely the original work of the authors.

\bibliographystyle{ACM-Reference-Format}
\bibliography{mcmfh-ref}

@String{Computing = "Computing" }

@String{Computer = "{IEEE} Computer" }

@String{Springer = "Springer-Verlag" }

@article{zhang2025biomedclipmultimodalbiomedicalfoundation,
  title={Biomedclip: a multimodal biomedical foundation model pretrained from fifteen million scientific image-text pairs},
  author={Zhang, Sheng and Xu, Yanbo and Usuyama, Naoto and Xu, Hanwen and Bagga, Jaspreet and Tinn, Robert and Preston, Sam and Rao, Rajesh and Wei, Mu and Valluri, Naveen and others},
  journal={arXiv preprint arXiv:2303.00915},
  year={2023}
}

@article{fedus2022reviewsparseexpertmodels,
  title={A review of sparse expert models in deep learning},
  author={Fedus, William and Dean, Jeff and Zoph, Barret},
  journal={arXiv preprint arXiv:2209.01667},
  year={2022}
}

@article{vaswani2023attentionneed,
  title={Attention is all you need},
  author={Vaswani, Ashish and Shazeer, Noam and Parmar, Niki and Uszkoreit, Jakob and Jones, Llion and Gomez, Aidan N and Kaiser, {\L}ukasz and Polosukhin, Illia},
  journal={Advances in neural information processing systems},
  volume={30},
  year={2017}
}

@article{fedus2022switchtransformersscalingtrillion,
  title={Switch transformers: Scaling to trillion parameter models with simple and efficient sparsity},
  author={Fedus, William and Zoph, Barret and Shazeer, Noam},
  journal={Journal of Machine Learning Research},
  volume={23},
  number={120},
  pages={1--39},
  year={2022}
}

@article{shazeer2017outrageouslylargeneuralnetworks,
  title={Outrageously large neural networks: The sparsely-gated mixture-of-experts layer},
  author={Shazeer, Noam and Mirhoseini, Azalia and Maziarz, Krzysztof and Davis, Andy and Le, Quoc and Hinton, Geoffrey and Dean, Jeff},
  journal={arXiv preprint arXiv:1701.06538},
  year={2017}
}

@article{esteva2019guide,
  title={A guide to deep learning in healthcare},
  author={Esteva, Andre and Robicquet, Alexandre and Ramsundar, Bharath and Kuleshov, Volodymyr and DePristo, Mark and Chou, Katherine and Cui, Claire and Corrado, Greg and Thrun, Sebastian and Dean, Jeff},
  journal={Nature medicine},
  volume={25},
  number={1},
  pages={24--29},
  year={2019},
  publisher={Nature Publishing Group US New York}
}

@article{acosta2022multimodal,
  title={Multimodal biomedical AI},
  author={Acosta, Juli{\'a}n N and Falcone, Guido J and Rajpurkar, Pranav and Topol, Eric J},
  journal={Nature medicine},
  volume={28},
  number={9},
  pages={1773--1784},
  year={2022},
  publisher={Nature Publishing Group US New York}
}

@article{wang2017survey,
  title={A survey on learning to hash},
  author={Wang, Jingdong and Zhang, Ting and Sebe, Nicu and Shen, Heng Tao and others},
  journal={IEEE transactions on pattern analysis and machine intelligence},
  volume={40},
  number={4},
  pages={769--790},
  year={2017},
  publisher={IEEE}
}

@inproceedings{jiang2017deep,
  title={Deep cross-modal hashing},
  author={Jiang, Qing-Yuan and Li, Wu-Jun},
  booktitle={Proceedings of the IEEE conference on computer vision and pattern recognition},
  pages={3232--3240},
  year={2017}
}

@article{wang2025cross,
  title={Cross-modal retrieval: a systematic review of methods and future directions},
  author={Wang, Tianshi and Li, Fengling and Zhu, Lei and Li, Jingjing and Zhang, Zheng and Shen, Heng Tao},
  journal={Proceedings of the IEEE},
  year={2025},
  publisher={IEEE}
}

@inproceedings{zhang2018deep,
  title={Deep cross-modal projection learning for image-text matching},
  author={Zhang, Ying and Lu, Huchuan},
  booktitle={Proceedings of the European conference on computer vision (ECCV)},
  pages={686--701},
  year={2018}
}

@inproceedings{zhen2019deep,
  title={Deep supervised cross-modal retrieval},
  author={Zhen, Liangli and Hu, Peng and Wang, Xu and Peng, Dezhong},
  booktitle={Proceedings of the IEEE/CVF conference on computer vision and pattern recognition},
  pages={10394--10403},
  year={2019}
}

@article{wu2020modality,
  title={Modality-specific and shared generative adversarial network for cross-modal retrieval},
  author={Wu, Fei and Jing, Xiao-Yuan and Wu, Zhiyong and Ji, Yimu and Dong, Xiwei and Luo, Xiaokai and Huang, Qinghua and Wang, Ruchuan},
  journal={Pattern Recognition},
  volume={104},
  pages={107335},
  year={2020},
  publisher={Elsevier}
}

@article{qin2025deep,
  title={Deep Discriminative Boundary Hashing for Cross-modal Retrieval},
  author={Qin, Qibing and Huo, Yadong and Zhang, Wenfeng and Huang, Lei and Nie, Jie},
  journal={IEEE Transactions on Circuits and Systems for Video Technology},
  year={2025},
  publisher={IEEE}
}

@article{zhu2025deep,
  title={Deep neighbor-coherence hashing with discriminative sample mining for supervised cross-modal retrieval},
  author={Zhu, Congcong and Qin, Qibing and Zhang, Wenfeng and Huang, Lei},
  journal={Expert Systems with Applications},
  volume={279},
  pages={127365},
  year={2025},
  publisher={Elsevier}
}

@article{huo2024deep2,
  title={Deep neighborhood-aware proxy hashing with uniform distribution constraint for cross-modal retrieval},
  author={Huo, Yadong and Qibing, Qin and Dai, Jiangyan and Zhang, Wenfeng and Huang, Lei and Wang, Chengduan},
  journal={ACM Transactions on Multimedia Computing, Communications and Applications},
  volume={20},
  number={6},
  pages={1--23},
  year={2024},
  publisher={ACM New York, NY}
}

@article{qin2025deep2,
  title={Deep Semantic-consistent Penalizing Hashing for Cross-modal Retrieval},
  author={Qin, Qibing and Wu, Lei and Zhang, Wenfeng and Huang, Lei and Nie, Jie},
  journal={IEEE Transactions on Multimedia},
  year={2025},
  publisher={IEEE}
}

@article{huo2024deep,
  title={Deep hierarchy-aware proxy hashing with self-paced learning for cross-modal retrieval},
  author={Huo, Yadong and Qin, Qibing and Zhang, Wenfeng and Huang, Lei and Nie, Jie},
  journal={IEEE Transactions on Knowledge and Data Engineering},
  year={2024},
  publisher={IEEE}
}

@article{tu2025cross,
  title={Cross-modal Hashing via Diverse Instances Matching},
  author={Tu, Junfeng and Liu, Xueliang and Huang, Zhen and Hao, Yanbin and Hong, Richang and Wang, Meng},
  journal={IEEE Transactions on Image Processing},
  year={2025},
  publisher={IEEE}
}

@article{tu2024two,
  title={Two-step discrete hashing for cross-modal retrieval},
  author={Tu, Junfeng and Liu, Xueliang and Hao, Yanbin and Hong, Richang and Wang, Meng},
  journal={IEEE Transactions on Multimedia},
  year={2024},
  publisher={IEEE}
}

@article{huo2023deep,
  title={Deep semantic-aware proxy hashing for multi-label cross-modal retrieval},
  author={Huo, Yadong and Qin, Qibing and Dai, Jiangyan and Wang, Lei and Zhang, Wenfeng and Huang, Lei and Wang, Chengduan},
  journal={IEEE Transactions on Circuits and Systems for Video Technology},
  volume={34},
  number={1},
  pages={576--589},
  year={2023},
  publisher={IEEE}
}

@inproceedings{liu2023multi,
  title={Multi-granularity interactive transformer hashing for cross-modal retrieval},
  author={Liu, Yishu and Wu, Qingpeng and Zhang, Zheng and Zhang, Jingyi and Lu, Guangming},
  booktitle={Proceedings of the 31st ACM International Conference on Multimedia},
  pages={893--902},
  year={2023}
}

@inproceedings{tu2022differentiable,
  title={Differentiable cross-modal hashing via multimodal transformers},
  author={Tu, Junfeng and Liu, Xueliang and Lin, Zongxiang and Hong, Richang and Wang, Meng},
  booktitle={Proceedings of the 30th ACM International Conference on Multimedia},
  pages={453--461},
  year={2022}
}

@inproceedings{lu2019vilbert,
  title={ViLBERT: Pretraining Task-Agnostic Visiolinguistic Representations for Vision-and-Language Tasks},
  author={Lu, Jiasen and Batra, Dhruv and Parikh, Devi and Lee, Stefan},
  booktitle={Advances in Neural Information Processing Systems (NeurIPS)},
  year={2019}
}

@inproceedings{tan2019lxmert,
  title={LXMERT: Learning Cross-Modality Encoder Representations from Transformers},
  author={Tan, Hao and Bansal, Mohit},
  booktitle={Proceedings of the 2019 Conference on Empirical Methods in Natural Language Processing (EMNLP)},
  year={2019}
}

@inproceedings{chen2020uniter,
  title={UNITER: UNiversal Image-TExt Representation Learning},
  author={Chen, Yen-Chun and Li, Linjie and Yu, Licheng and Kholy, Ahmed El and Ahmed, Faisal and Gan, Zhe and Cheng, Yu and Liu, Jingjing},
  booktitle={European Conference on Computer Vision (ECCV)},
  year={2020}
}

@inproceedings{kiela2019mmbt,
  title={Supervised Multimodal Bitransformers for Classifying Images and Text},
  author={Kiela, Douwe and Bulian, Jannis and Baziotis, Christos and Firooz, Hamed and Giorgi, John and Jin, Zhiguo and Mohan, Arjun and et al.},
  booktitle={Proceedings of the 2019 Conference on Empirical Methods in Natural Language Processing (EMNLP)},
  year={2019}
}

@inproceedings{tsai2019multimodal,
  title={Multimodal Transformer for Unaligned Multimodal Language Sequences},
  author={Tsai, Yao-Hung Hubert and Bai, Shaojie and Yamada, Paul Pu Liang and Morency, Louis-Philippe and Salakhutdinov, Ruslan},
  booktitle={Proceedings of the Annual Meeting of the Association for Computational Linguistics (ACL)},
  year={2019}
}

@inproceedings{shazeer2017outrageously,
  title={Outrageously Large Neural Networks: The Sparsely-Gated Mixture-of-Experts Layer},
  author={Shazeer, Noam and Mirhoseini, Azalia and Maziarz, Krzysztof and Davis, Andy and Le, Quoc V and Hinton, Geoffrey and Dean, Jeff},
  booktitle={International Conference on Learning Representations (ICLR)},
  year={2017}
}

@article{srivastava2014dropout,
  title={Dropout: A Simple Way to Prevent Neural Networks from Overfitting},
  author={Srivastava, Nitish and Hinton, Geoffrey and Krizhevsky, Alex and Sutskever, Ilya and Salakhutdinov, Ruslan},
  journal={Journal of Machine Learning Research},
  volume={15},
  number={1},
  pages={1929--1958},
  year={2014}
}

@article{XIA2023101968,
title = {When CLIP meets cross-modal hashing retrieval: A new strong baseline},
journal = {Information Fusion},
volume = {100},
pages = {101968},
year = {2023},
issn = {1566-2535},
doi = {https://doi.org/10.1016/j.inffus.2023.101968},
url = {https://www.sciencedirect.com/science/article/pii/S1566253523002841},
author = {Xinyu Xia and Guohua Dong and Fengling Li and Lei Zhu and Xiaomin Ying}
}

@ARTICLE{10855579,
  author={Qin, Qibing and Wu, Lei and Zhang, Wenfeng and Huang, Lei and Nie, Jie},
  journal={IEEE Transactions on Multimedia}, 
  title={Deep Semantic-consistent Penalizing Hashing for Cross-modal Retrieval}, 
  year={2025},
  volume={},
  number={},
  pages={1-14},
  doi={10.1109/TMM.2025.3535306}}

@ARTICLE{11003934,
  author={Qin, Qibing and Huo, Yadong and Zhang, Wenfeng and Huang, Lei and Nie, Jie},
  journal={IEEE Transactions on Circuits and Systems for Video Technology}, 
  title={Deep Discriminative Boundary Hashing for Cross-modal Retrieval}, 
  year={2025},
  volume={},
  number={},
  pages={1-1},
  doi={10.1109/TCSVT.2025.3570128}}

@article{kingma2017adammethodstochasticoptimization,
  title={Adam: A method for stochastic optimization},
  author={Kingma, Diederik P},
  journal={arXiv preprint arXiv:1412.6980},
  year={2014}
}

@inproceedings{kim2023improving,
  title={Improving cross-modal retrieval with set of diverse embeddings},
  author={Kim, Dongwon and Kim, Namyup and Kwak, Suha},
  booktitle={Proceedings of the IEEE/CVF conference on computer vision and pattern recognition},
  pages={23422--23431},
  year={2023}
}

@inproceedings{zhang2024fine,
  title={Fine-grained Prototypical Voting with Heterogeneous Mixup for Semi-supervised 2D-3D Cross-modal Retrieval},
  author={Zhang, Fan and Hua, Xian-Sheng and Chen, Chong and Luo, Xiao},
  booktitle={Proceedings of the IEEE/CVF Conference on Computer Vision and Pattern Recognition},
  pages={17016--17026},
  year={2024}
}

@article{bodenreider2004unified,
  title     = {The Unified Medical Language System (UMLS): integrating biomedical terminology},
  author    = {Bodenreider, Olivier},
  journal   = {Nucleic Acids Research},
  volume    = {32},
  number    = {suppl\_1},
  pages     = {D267--D270},
  year      = {2004},
  publisher = {Oxford University Press},
  doi       = {10.1093/nar/gkh061}
}

@misc{openi_dataset,
  title        = {Open-I: An Open Access Biomedical Image Search Engine},
  author       = {{U.S. National Library of Medicine}},
  year         = {2013},
  howpublished = {\url{https://openi.nlm.nih.gov}},
  note         = {Accessed: 2025-06-07}
}

@inproceedings{pelka2018roco,
  title     = {Radiology Objects in COntext (ROCO): A Multimodal Image Dataset},
  author    = {Pelka, Obioma and Koitka, Sven and Rückert, Johannes and Nensa, Felix and Friedrich, Christoph M.},
  booktitle = {Intravascular Imaging and Computer Assisted Stenting and Large-Scale Annotation of Biomedical Data and Expert Label Synthesis},
  series    = {Lecture Notes in Computer Science},
  volume    = {11043},
  pages     = {180--189},
  publisher = {Springer},
  year      = {2018},
  doi       = {10.1007/978-3-030-01364-6_20}
}

\end{document}